# Research Output on Alopecia Areata Disease: A Scientometric Analysis of Publications from 2010 to 2019

Muneer Ahmad
muneerbangroo@gmail.com

Dr. M.Sadik Batcha
*Annamalai University*

# Research Output on Alopecia Areata Disease: A Scientometric Analysis of Publications from 2010 to 2019

Muneer Ahmad[1] Dr. M Sadik Batcha[2]

[1]*Research Scholar, Department of Library and Information Science, Annamalai University, Annamalai nagar, muneerbangroo@gmail.com*
[2]*Research Supervisor & Mentor, Professor and University Librarian, Annamalai University, Annamalai nagar, msbau@rediffmail.com*

**Abstract**

The present study is undertaken to find out the publication trends on Alopecia Areata Disease during 2010-2019 from the global perspective. The study mainly focus on distribution of research output, top journals for publications, most prolific authors, authorship pattern, and citations pattern on Alopecia Areata Disease. The results indicate that highest growth rate of publications occurred during the year 2019. Columbia University topped the scene among all institutes. The maximum publications were more than four authored publications. Christiano AM and Clynes R were found to be the most prolific authors. It is also found that most of the prolific authors (by number of publications) do appear in highly cited publications' list. Alopecia Areata Disease researchers mostly preferred using article publications to communicate their findings.

## 1. Introduction

Alopecia areata is an autoimmune disorder characterized by transient, non-scarring hair loss and preservation of the hair follicle. Hair loss can acquire numerous forms ranging from loss in well-defined patches to diffuse or total hair loss, which can affect all hair bearing sites. Patchy alopecia disturbing the scalp is the most widespread type. Alopecia areata affects nearly 2% of the general population at some point during their life span. Skin biopsies of alopecia areata affected skin confirm a lymphocytic infiltrate in and around the bulb or the lower part of the hair follicle in anagen (hair growth) phase. A breakdown of immune privilege of the hair follicle is reflection to be an important driver of alopecia areata. Genetic studies in patients and mouse models showed that alopecia areata is a multifaceted, polygenic disease. Several genetic receptiveness loci were identified related with signaling pathways that are significant to hair follicle cycling and development. Alopecia areata is frequently diagnosed based on clinical manifestations, but dermoscopy and histopathology can be supportive. Alopecia areata is complicated to control medically, but recent advances in understanding the molecular mechanisms have bare new treatments and the opportunity of diminution in the near future (Pratt, King, Messenger, Christiano, & Sundberg, 2017).

## 2. Review of Literature

Review of related literature, is a primary component of any research investigation, enables the investigator to understand the earlier research interests, research patterns and the magnitude of the research output in a field of knowledge. Hence, there is a need to review such significant works and their relevance for the present study. (Batcha & Ahmad, 2017) obtained the analysis of two journals Indian Journal of Information Sources and Services (IJSS) which is of Indian origin and Pakistan Journal of Library and Information Science (PJLIS) from Pakistan origin and studied them comparatively with scientometric indicators like year wise distribution of articles, pattern of authorship and productivity, degree of collaboration, pattern of co-authorship, average length of papers, average keywords. The collaboration with foreign authors of both the countries is negligible (1.37% of articles) from India and (4.10% of articles) from Pakistan.

(Ahmad, Batcha, Wani, Khan, & Jahina, 2018) studied Webology journal one of the reputed journals from Iran was explored through scientometric analysis. The study aims to provide a comprehensive analysis regarding the journal like year wise growth of research articles, authorship pattern, author productivity, and subjects taken by the authors over the period of 5 years from 2013 to 2017. The findings indicate that 62 papers were published in the journal during the study period. The articles having collaborative nature were high in number. Regarding the subject concentration of papers of the journal, Social Networking, Web 2.0, Library 2.0 and Scientometrics or Bibliometrics were highly noted.

(Batcha, Jahina, & Ahmad, 2018) has examined the DESIDOC Journal by means of various scientometric indicators like year wise growth of research papers , authorship pattern, subjects and themes of the articles over the period of five years from 2013 to 2017. The study reveals that 227 articles were published over the five years from 2013 to 2017. The authorship pattern was highly collaborative in nature.  The maximum numbers of articles (65 %) have ranged their thought contents between 6 and 10 pages.

(Ahmad & Batcha, 2019) analyzed research productivity in Journal of Documentation (JDoc) for a period of 30 years between 1989 and 2018. Web of Science a service from Clarivate Analytics has been consulted to obtain bibliographical data and it has been analysed through Bibexcel and Histcite tools to present the datasets. Analysis part deals with local and global citation level impact, highly prolific authors and their research output, ranking of prominent institution and countries. In addition to this scientographical mapping of bibliographical data is obtainable through VOSviewer, which is open source mapping software.

(Ahmad & Batcha, in 2019) studied the scholarly communication of Bharathiar University which is one of the vibrant universities in Tamil Nadu. The study find out the impact of research produced, year-wise research output, citation impact at local and global level, prominent authors and their total output, top journals of publications, top collaborating countries which collaborate with the university authors, highly industrious departments and trends in publication of the university during 2009 through 2018. In addition the study used scientographical mapping of data and presented it through graphs using VOSviewer software mapping technique.

(Ahmad, Batcha, & Jahina, 2019) quantitatively measured the research productivity in the area of artificial intelligence at global level over the study period of ten years (2008-2017). The study acknowledged the trends and features of growth and collaboration pattern of artificial intelligence research output. Average growth rate of artificial intelligence per year increases at the rate of 0.862. The multi-authorship pattern in the study is found high and the average number of authors per paper is 3.31. Collaborative Index is noted to be the highest range in the year 2014 with 3.50. Mean CI during the period of study is 3.24. This is also supported by the mean degree of collaboration at the percentage of 0.83 .The mean CC observed is 0.4635. Regarding the application of Lotka's Law of authorship productivity in the artificial intelligence literature it proved to be fit for the study.

(Batcha, Dar, & Ahmad, 2019) presented a scientometric analysis of the journal titled "Cognition" for a period of 20 years from 1999 to 2018. The present study was conducted with an aim to provide a summary of research activity in current journal and characterize its most aspects. The research coverage includes the year wise distribution of articles, authors, institutions, countries and citation analysis of the journal. The analysis showed that 2870 papers were published in journal of Cognition from 1999 to 2018. The study identified top 20 prolific authors, institutions and countries of the journal. Researchers from USA have made the most percentage of contributions.

(Batcha, Dar, & Ahmad, 2020) conducts a scientometric study of the *Modern Language Journal* literature from 1999 to 2018. A total of 2564 items resulted from the publication name using "Modern Language Journal" as the search term was retrieved from the Web of Science Database. Based on the number of publications during the study period, no consistent growth was observed in the research activities pertaining to the journal. The annual distribution of publications, number of authors, institution productivity, country wise publications and Citations are analyzed. Highly productive authors, institutions, and countries are identified. The results reveal that the maximum number of papers 179 is published in the year 1999. It was also observed that Byrnes H is the most productive, contributed 51 publications and Kramsch C is most cited author in the field having 543 global citations. The highest number (38.26%) of publications, contributed from USA and the foremost productive establishment was University of Iowa.

(Ahmad, Batcha, & Dar, 2020) studied the Brain and Language journal which is an interdisciplinary journal, publishes articles that explicate the complex relationships among language, brain, and behavior and is one such journal which is concerned with investigating the neural correlates of Language. The study aims at mapping the structure of the *Brain and Language* journal. The journal looks into the intrinsic relationship between language and brain. The study demonstrates and elaborates on the various aspects of the Journal, such as its chronology wise total papers, most productive authors, citations, average citation per paper, institution and country wise distribution of publications for a period of 20 years.

(Ahmad & Batcha, 2020) explores and analyses the trend of world literature on "Coronavirus Disease" in terms of the output of research publications as indexed in the Science Citation Index Expanded (SCI-E) of Web of Science during the period from 2011 to 2020. The study found that 6071 research records have been published on Coronavirus Disease. The various scientometric components of the research records published in the study period were studied. The study reveals the various aspects of Coronavirus Disease literature such as year wise distribution, relative growth rate, doubling time of literature, geographical wise, organization wise, language wise, form wise , most prolific authors, and source wise.

(Ahmad & Batcha, 2020) analyzed the application of Lotka's law to the research publication, in the field of Dyslexia disease. The data related to Dyslexia were extracted from web of science database, which is a scientific, citation and indexing service, maintained by Clarivate Analytics. A total of 5182 research publications were published by the researchers, in the field of Dyslexia. The study found out that, the Lotka's inverse square law is not fit for this data. The study also analyzed the authorship pattern, Collaborative Index (CI), Degree of Collaboration (DC), Co-authorship Index (CAI), Collaborative Co-efficient (CC), Modified Collaborative Co-efficient (MCC), Lotka's Exponent value, Kolmogorov-Smirnov Test (K-S Test), Relative Growth Rate and Doubling Time.

(Umar, Ahmad, & Batcha, 2020) studied and focused on the growth and development of Library and Culture research in forms of publications reflected in Web of Science database, during the span of 2010-2019. A total 890 publications were found and the highest 124 (13.93%) publications published in 2019.The analysis maps comprehensively the parameters of total output, growth of output, authorship, institution wise and country-level collaboration patterns, major contributors (individuals, top publication sources, institutions, and countries).

(Ahmad & Batcha, 2020) studied and examined 4698 Indian Coronary Artery Disease research publications, as indexed in Web of Science database during 1990-2019, with a view to understand their growth rate, global share, citation impact, international collaborative papers, distribution of publications by broad subjects, productivity and citation profile of top organizations and authors, and preferred media of communication.

(Jahina, Batcha, & Ahmad, 2020) study deals a scientometric analysis of 8486 bibliometric publications retrieved from the Web of Science database during the period 2008 to 2017. Data is collected and analyzed using Bibexcel software. The study focuses on various aspect of the quantitative research such as growth of papers (year wise), Collaborative Index (CI), Degree of Collaboration (DC), Co-authorship Index (CAI), Collaborative Co-efficient (CC), Modified Collaborative Co-Efficient (MCC), Lotka's Exponent value, Kolmogorov-Smirnov test (K-S Test).

## 3. Objectives

The main objective of the present study is to study the growth of research output in Alopecia Areata Disease globally. Moreover, the study has been performed:

- To find out the type of documents containing Alopecia Areata Disease research output in world during 2010-2019;

- To analyse the year wise distribution and growth of literature on Alopecia Areata Disease during 2010-2019;
- To identify the top institutions conducting research on Alopecia Areata Disease;
- To identify the most prolific authors conducting research Alopecia Areata Disease;
- To study the authorship pattern in Alopecia Areata Disease research;
- To study the top sources preferred by authors for publishing Alopecia Areata Disease research.

## 4. Methodology

The present study is a scientometric analysis of Alopecia Areata Disease research publications. A total of 2127 records have been extracted from the Web of Science database in the '.txt' format covering the period (2010-2019). The search string used for data extraction is: "Alopecia Areata" This search has been refined to limit the period from 2010 to 2019. Data filtering has been performed manually to remove irrelevant record entries. Bibliometrix Package in RStudio has been used for analyzing the data and it has also been used for tabulation and visualization of Results.

Calculations and statistical techniques were applied in the excel sheet to draw specific results. Total Publications (TP), Total Citations (TC), Average Citations per Paper (ACPP) h-index, g-index and m-index was calculated during analysis. ACPP is calculated by dividing the total citations received by the number of papers. The h-index was suggested by Jorge H. Hirch in 2005 (Hirsch, 2010). A scientist/ journal/ institution has index h if its h papers have atleast h citations each. Egghe defines g-index as "the highest rank such that the top g papers have, together, at least g2 citations. This also means that the top g + 1 have less than (g + 1)2 papers". The g-index is always higher or equal to h-index, as has been also stated by (Egghe, 2006). m-index is another variant of the h-index that displays h-index per year since first publication. The h-index tends to increase with career length, and m-index can be used in situations where this is a shortcoming, such as comparing researchers within a field but with very different career lengths. The m-index inherently assumes unbroken research activity since the first publication.

## 5. Data Analysis and Findings
### 5.1. Type of Publications

Different kind of publications in which research work on Alopecia Areata is contributed during last 10 years is listed in Table 1. Out of total publications 1032 (48.52 %) are articles, 446 (20.97 %) are meeting abstracts, 266 (12.51 %) are letters, 266 (12.51 %) are reviews, 80 (3.76 %) are editorial material, 9 (0.42 %) are article; early access, 9 (0.42 %) are corrections, 8 (0.38 %) are article; proceeding papers, 3 (0.14 %) are news items, and 3 (0.14 %) are review; early access , 2 (0.09 %) are letter; early access, 1 (0.05 ,%) is article; book chapter, 1 (0.05 %) is article; retracted publication and 1 (0.05 %) is biographical item . It is apparent that more research output was produced in the form of articles. It is also evident that in spite of more research output was produced in articles but ACPP of article; book chapter published was a fair amount (65.00) compared to articles (16.39). ACPP of review (22.18), news item (11), letter (9.51), article; retracted publication (9), article; early access (3.98). Other type of documents had ACPP less than 2.Thus; it was observed that article; book chapter, review and articles received more citations than other forms of documents.

### Table 1 : Publication Types

| Type | TP | % | TC | ACPP |
|---|---|---|---|---|
| Article | 1032 | 48.52 | 16912 | 16.39 |
| Meeting Abstract | 446 | 20.97 | 101 | 0.23 |
| Letter | 266 | 12.51 | 2530 | 9.51 |
| Review | 266 | 12.51 | 5899 | 22.18 |
| Editorial Material | 80 | 3.76 | 318 | 3.98 |
| Article; Early Access | 9 | 0.42 | 5 | 0.56 |
| Correction | 9 | 0.42 | 2 | 0.22 |
| Article; Proceedings Paper | 8 | 0.38 | 76 | 9.50 |
| News Item | 3 | 0.14 | 33 | 11.00 |
| Review; Early Access | 3 | 0.14 | 2 | 0.67 |
| Letter; Early Access | 2 | 0.09 | 0 | 0.00 |
| Article; Book Chapter | 1 | 0.05 | 65 | 65.00 |
| Article; Retracted Publication | 1 | 0.05 | 9 | 9.00 |
| Biographical-Item | 1 | 0.05 | 1 | 1.00 |

TP= "Total Publications", TC= "Total Citations", ACPP= "Average Citations per Paper"

**5.2. Distribution of Research Publications**

There has been increase in publications from 2010-2019. During last year, about 15.75 percent research output on Alopecia Areata was contributed in the year 2019. Table 2 shows the distribution of research output year wise. It is very apparent that highest growth rate occurs in the year 2013 (56.00 %) followed by 2019 (26.89 %). It is also evident from the table a decrease in research output during the year 2011 (-15.76), 2014 (-11.97) and 2015 (-8.25). There is a continuous increase in the research output from the year 2016-2019.

Table 2 : Distribution of Papers during 2010-2019

| Year | Articles | % of TP | CO | % of Growth |
|---|---|---|---|---|
| 2010 | 165 | 7.76 | - | - |
| 2011 | 139 | 6.54 | 304 | -15.76 |
| 2012 | 150 | 7.05 | 454 | 7.91 |
| 2013 | 234 | 11.00 | 688 | 56.00 |
| 2014 | 206 | 9.69 | 894 | -11.97 |
| 2015 | 189 | 8.89 | 1083 | -8.25 |
| 2016 | 204 | 9.59 | 1287 | 7.94 |
| 2017 | 241 | 11.33 | 1528 | 18.14 |
| 2018 | 264 | 12.41 | 1792 | 9.54 |
| 2019 | 335 | 15.75 | 2127 | 26.89 |
| Total | 2127 | 100.00 | - | - |

TP= "Total Publications", CO= "Cumulative Output", Formula of Growth= "Final Value-Start Value/Start Value X100"

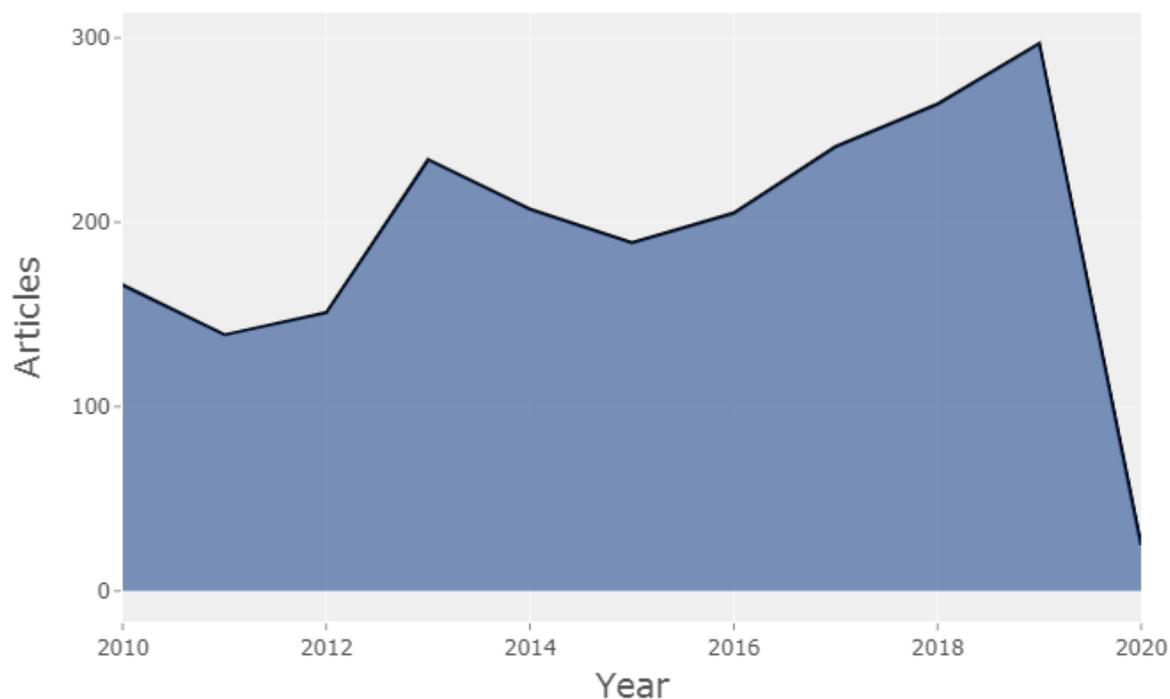

### 5.3. Institution-wise Research Share

The top 20 institutions that produced highest research outputs on Alopecia Areata during the period under study are listed in Table 3. Table 3 summarizes total articles, the total citation score, and average citation per paper of the publications of these institutions. In total, 1939 institutions, including 3439 subdivisions published 2127 research papers during 2010 – 2019. The topmost 20 institutions involved in this research have published 22 and more research articles. The mean average is 1.10 research articles per Institution. Out of 1939 institutions, top 20 institutions published 725 (37.39%) research papers and the rest of the institutions published 1214 (62.61%) research papers respectively. It is also observed that among twenty Institutions which contributed highest research output on Alopecia Areata are Colombia University, University Manchester and University Miami. Columbia University took the lead by producing 127 followed by University Manchester with 61 and University Miami with 55 research publications respectively. In terms of citations, Columbia University received highest citations i.e. 2622 for 127 total research publications. It is also noticed that Technion Israel Institution Technology had highest ACPP (38.22).

Table 3: Top Institutions Research Output

| S.No. | Institution | TP | % | TC | ACPP |
|---|---|---|---|---|---|
| 1 | Columbia University | 127 | | 2622 | 20.65 |
| 2 | University Manchester | 61 | | 1946 | 31.90 |
| 3 | University Miami | 55 | | 827 | 15.04 |
| 4 | University British Columbia | 46 | | 1363 | 29.63 |
| 5 | University Minnesota | 39 | | 905 | 23.21 |
| 6 | University Munster | 35 | | 790 | 22.57 |
| 7 | Yonsei University | 34 | | 190 | 5.59 |
| 8 | Jackson Lab | 32 | | 495 | 15.47 |
| 9 | University Colorado | 31 | | 1036 | 33.42 |
| 10 | Northwestern University | 29 | | 240 | 8.28 |
| 11 | Kyung Hee University | 26 | | 165 | 6.35 |
| 12 | Yale University | 26 | | 878 | 33.77 |
| 13 | Icahn Sch Med Mt Sinai | 24 | | 314 | 13.08 |
| 14 | University Texas MD Anderson Canc Ctr | 24 | | 889 | 37.04 |
| 15 | Hamamatsu University School Med | 23 | | 432 | 18.78 |
| 16 | Harvard University | 23 | | 708 | 30.78 |
| 17 | Technion Israel Inst Technol | 23 | | 879 | 38.22 |
| 18 | University Penn | 23 | | 799 | 34.74 |
| 19 | University Munich | 22 | | 400 | 18.18 |
| 20 | University Tehran Med Science | 22 | | 194 | 8.82 |

TP= "Total Publications", TC= "Total Citations", ACPP= "Average Citations per Paper".

## 5.4. Most Prolific Authors

The list of twenty top authors who produced highest contribution to research output on Alopecia Areata is given in Table 4. In terms of number of publications, Christiano AM is the most productive author with 84 publications followed by Clynes R 60, Paus R 53, and Jabbari A 52 publications. It is also noted that 4 out of 20 prolific authors contributed more than fifty research publications each while rest 16 authors contributed less than 50 publications each. The ACPP on research output contributed by Paus R (24.57) was recorded highest that was nearly followed by Petukhova L (22.32). The h index is highest for Christiano AM (17) and Paus R (17) respectively followed by Mackay Wiggan J (13) Tostia A (13) respectively. The data set puts forth that the authors Christiani AM with 39 g index, Paus R with 36 g index, Clynes R with 29 g index and Petukhova L with 29 g index. Christiano AM (1.55), Paus R (1.55), Jabbari A (1.11), and Clynes R (1.10) are having the highest m index respectively.

Table 4: Most Prolific Authors

| Rank | Author | TP | TC | ACPP | h index | g index | m index |
|---|---|---|---|---|---|---|---|
| 1 | Christiano AM | 84 | 1527 | 18.18 | 17 | 39 | 1.55 |
| 2 | Clynes R | 60 | 847 | 14.12 | 11 | 29 | 1.10 |
| 3 | Paus R | 53 | 1302 | 24.57 | 17 | 36 | 1.55 |
| 4 | Jabbari A | 52 | 819 | 15.75 | 10 | 28 | 1.11 |
| 5 | Christiano A | 41 | 11 | 0.27 | 2 | 3 | 0.25 |
| 6 | Petukhova L | 38 | 848 | 22.32 | 7 | 29 | 0.64 |
| 7 | Dai Z | 36 | 7 | 0.19 | 1 | 1 | 0.11 |
| 8 | Mackay Wiggan J | 33 | 686 | 20.79 | 13 | 26 | 1.09 |
| 9 | Duvic M | 32 | 588 | 18.38 | 7 | 24 | 0.64 |
| 10 | Mcelwee KJ | 31 | 668 | 21.55 | 12 | 25 | 1.09 |
| 11 | Shapiro J | 31 | 666 | 21.48 | 12 | 25 | 1.09 |
| 12 | Tosti A | 31 | 517 | 16.68 | 13 | 22 | 1.30 |
| 13 | Hordinsky M | 30 | 538 | 17.93 | 6 | 23 | 0.55 |
| 14 | De Jong A | 29 | 459 | 15.83 | 5 | 21 | 0.56 |
| 15 | Gilhar A | 29 | 543 | 18.72 | 11 | 23 | 1.00 |
| 16 | Bertolini M | 27 | 354 | 13.11 | 7 | 18 | 0.64 |
| 17 | Ito T | 27 | 240 | 8.89 | 8 | 15 | 0.73 |
| 18 | Lee WS | 27 | 91 | 3.37 | 6 | 8 | 0.75 |
| 19 | Sundberg JP | 27 | 306 | 11.33 | 9 | 17 | 0.82 |
| 20 | Sim WY | 26 | 99 | 3.81 | 6 | 9 | 0.55 |

TP= "Total Publications", TC= "Total Citations", ACPP= "Average Citations per Paper".

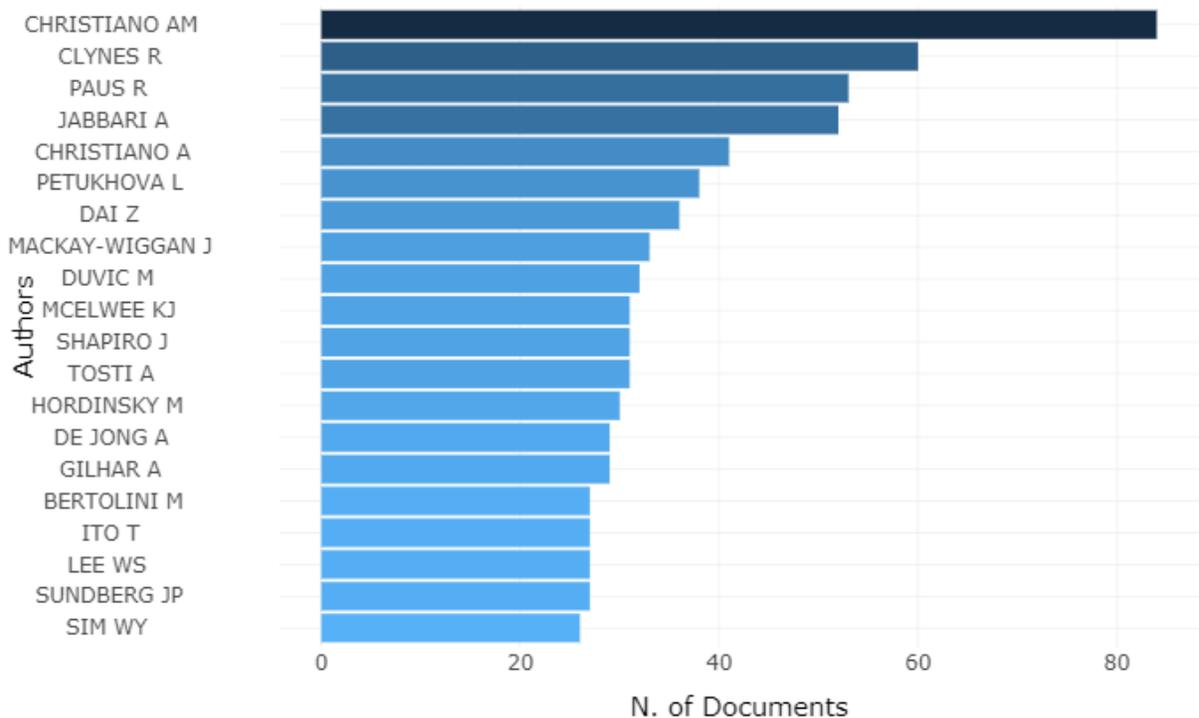

## 5.5. Authorship Pattern

Table 5 illustrates the overall and year wise distribution of authorship trend. It is evident from the Table 5 that only 7.62 per cent publications were single authored publications while rest of 92.38 had two or more authors. The maximum number of publications were four authored (16.46 %) nearly followed by three authored publications (16.27 %), two authored (14.86 %), five authored (12.69%), six authored (9.69 %) and one authored (7.62 %) publications. Seven to ten authored publications accounted for 16.64 per cent only while more than 10 authored publications accounted for 5.78%.

Table 5: Authorship Pattern

| Author(s) | Total Research Output in Ten Years | | | | | | | | | | Total Research Output | |
|---|---|---|---|---|---|---|---|---|---|---|---|---|
| | 2010 | 2011 | 2012 | 2013 | 2014 | 2015 | 2016 | 2017 | 2018 | 2019 | Total | % |
| Single | 15 | 23 | 10 | 25 | 18 | 10 | 14 | 12 | 17 | 18 | 162 | 7.62 |
| Two Author | 20 | 15 | 28 | 31 | 33 | 26 | 35 | 42 | 47 | 39 | 316 | 14.86 |
| Three | 29 | 30 | 18 | 37 | 30 | 29 | 27 | 47 | 44 | 55 | 346 | 16.27 |
| Four | 32 | 24 | 20 | 32 | 34 | 28 | 29 | 46 | 42 | 63 | 350 | 16.46 |
| Five | 23 | 13 | 17 | 24 | 23 | 24 | 24 | 34 | 40 | 48 | 270 | 12.69 |
| Six | 20 | 14 | 19 | 27 | 18 | 21 | 24 | 17 | 16 | 30 | 206 | 9.69 |
| Seven | 7 | 10 | 15 | 17 | 19 | 12 | 10 | 11 | 23 | 23 | 147 | 6.91 |
| Eight | 9 | 2 | 12 | 15 | 6 | 13 | 15 | 10 | 8 | 16 | 106 | 4.98 |
| Nine | 2 | | 1 | 8 | 9 | 6 | 7 | 5 | 6 | 13 | 57 | 2.68 |
| Ten | 2 | 3 | 3 | 8 | 2 | 4 | 10 | 2 | 3 | 7 | 44 | 2.07 |
| More than 10 | 7 | 5 | 8 | 10 | 15 | 16 | 10 | 15 | 17 | 20 | 123 | 5.78 |
| Total | 166 | 139 | 151 | 234 | 207 | 189 | 205 | 241 | 263 | 332 | 2127 | 100.00 |
| % | 7.80 | 6.54 | 7.10 | 11.00 | 9.73 | 8.89 | 9.64 | 11.33 | 12.36 | 15.61 | 100.00 | |

### 5.6. Top Journals Preferred for Publication

The total number of 2127 publications on Alopecia Areata from 2010 to 2019 appeared in 441 different sources. The top 20 journals preferred for publications on Alopecia Areata are listed in Table 6 which accounted for 54.77 per cent of total research publications during the period under study. Journal *of Investigative Dermatology has* published highest (265) publications on Alopecia Areata followed by Journal of the American Academy of Dermatology (171). According to the journals preferred for publication output from the table 6 the journal wise distribution of research documents, Journal of Investigative Dermatology has the highest number of research documents 265 with 1136 of total citation score and 17, 33 and 1.55 h index, g index and m index respectively being prominent among the 20 journals and it stood in first rank position. Journal of the American Academy of Dermatology has 171 research documents and it stood in second position with 2564 of total citation score and 27, 47and 2.45 h index, g index and m index score were scaled. It is followed by Journal of Dermatology with 87 records and it stood in third rank position along with 531 of total citation score and 12, 20 and 1.09 h, g, and m index score measured.

**Table 6: Top 20 Journals for Publications**

| S.No. | Source of Publication | NP | TC | ACPP | h index | g index | m index |
|---|---|---|---|---|---|---|---|
| 1 | Journal of Investigative Dermatology | 265 | 1136 | 4.29 | 17 | 33 | 1.55 |
| 2 | Journal of The American Academy of Dermatology | 171 | 2564 | 14.99 | 27 | 47 | 2.45 |
| 3 | Journal of Dermatology | 87 | 531 | 6.10 | 12 | 20 | 1.09 |
| 4 | British Journal of Dermatology | 78 | 883 | 11.32 | 19 | 28 | 1.11 |
| 5 | Experimental Dermatology | 74 | 447 | 6.04 | 12 | 20 | 1.09 |
| 6 | Journal of The European Academy of Dermatology And Venereology | 63 | 683 | 10.84 | 16 | 22 | 1.45 |
| 7 | International Journal of Dermatology | 57 | 384 | 6.74 | 12 | 16 | 1.09 |
| 8 | Australasian Journal of Dermatology | 41 | 98 | 2.39 | 6 | 9 | 0.60 |
| 9 | Dermatologic Therapy | 40 | 341 | 8.53 | 11 | 17 | 1.00 |
| 10 | Annals of Dermatology | 31 | 227 | 7.32 | 9 | 14 | 0.90 |

| 11 | Acta Dermato-Venereologica | 29 | 279 | 9.62 | 9 | 16 | 0.82 |
| 12 | European Journal of Dermatology | 29 | 175 | 6.03 | 7 | 12 | 0.64 |
| 13 | Pediatric Dermatology | 28 | 179 | 6.39 | 6 | 12 | 0.55 |
| 14 | Clinical And Experimental Dermatology | 27 | 245 | 9.07 | 11 | 14 | 1.00 |
| 15 | Jama Dermatology | 26 | 566 | 21.77 | 12 | 23 | 1.50 |
| 16 | Journal of Drugs In Dermatology | 26 | 212 | 8.15 | 8 | 13 | 0.73 |
| 17 | Journal of Dermatological Treatment | 25 | 166 | 6.64 | 7 | 12 | 0.69 |
| 18 | Indian Journal Of Dermatology | 24 | 93 | 3.88 | 5 | 9 | 0.71 |
| 19 | Dermatology | 22 | 238 | 10.82 | 10 | 14 | 0.91 |
| 20 | Journal Der Deutschen Dermatologischen Gesellschaft | 22 | 55 | 2.50 | 5 | 7 | 0.62 |

NP= "Number of Publications", TC= "Total Citations", ACPP= "Average Citations per Paper".

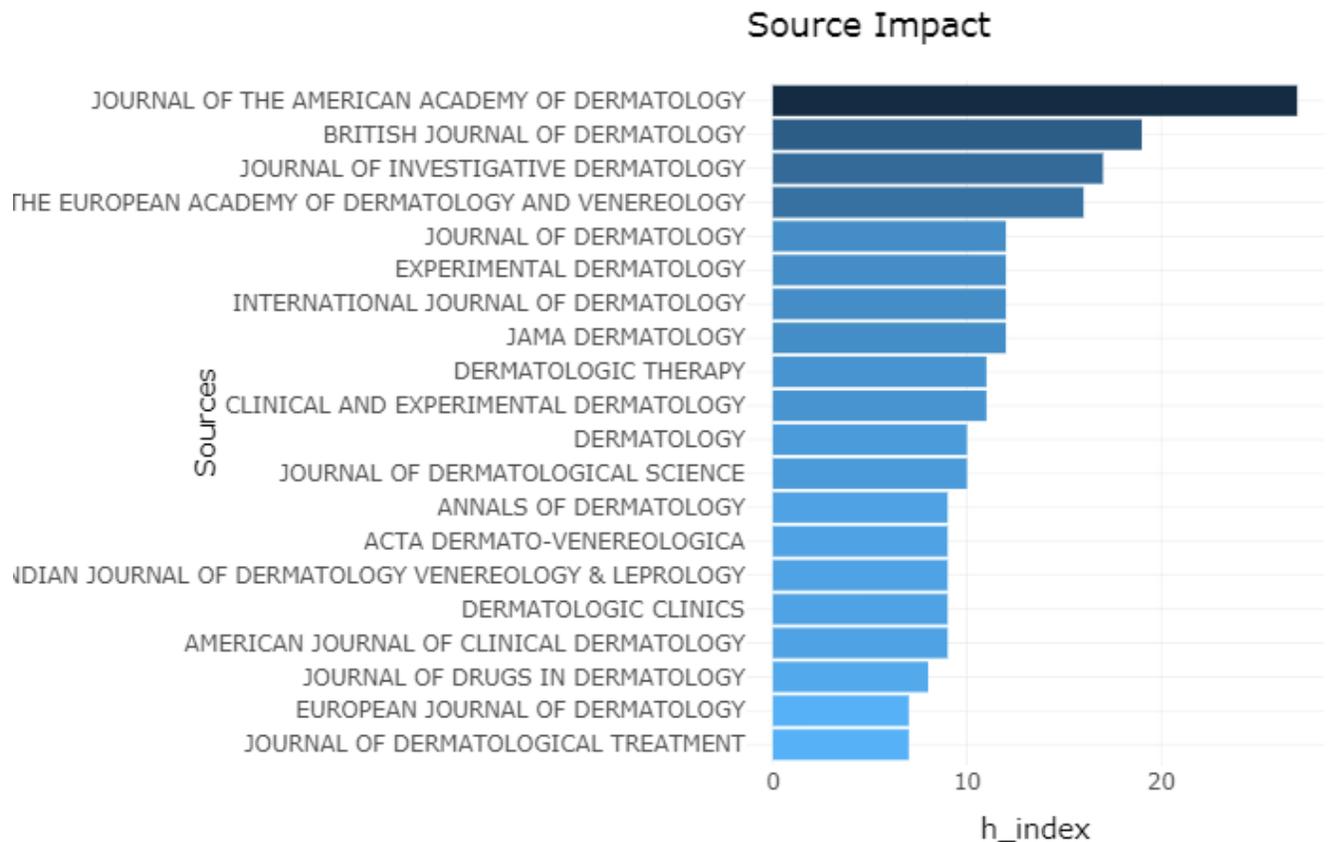

## 6. Conclusion

The study explores the 10 years research output on Alopecia Areata. It was found that a total number of 2127 papers on Alopecia Areata were published during 2010-2019 which received 25953 citations with ACPP of 12.20. Growth rate was highest (56%) in the year 2013. ACCP of Technion Israel Institute Technology was highest with 38.22 average citations per paper. Nearly 54.77 per cent of research on Alopecia Areata was published in 20 journals among which Journal of Investigative Dermatology produced highest research output on Alopecia Areata. Christiano

AM and Clynes R were the front runners in terms of number of publications but in terms of citations and ACPP Christiano A M and Paus R remained at top. Only 7.62 per cent publications were single authored publications while rest of 92.38 had two or more authors. Among all type of publications, reviews and articles received more citations. The study depicts that research work on Alopecia Areata was very less in earlier years but increased during the later years of study. Major research output on Alopecia Areata was produced in the year 2019 followed by 2018.